\UseRawInputEncoding
\documentclass[a4paper,twosides]{article}
\usepackage{CJK,multicol,multirow,graphicx,fancyhdr,epstopdf}
\usepackage{amsmath,amsfonts,amssymb,bm,upgreek,mathrsfs,ccmap,mathcomp}
\usepackage[pagewise,switch,columnwise]{lineno}
\usepackage[compress,nospace]{cite}
\usepackage[dvipsnames]{xcolor}
\usepackage{CPL-2023}
\usepackage{orcidlink}
\newcommand{\cplyear}{2026} \newcommand{\cplvol}{xx}
 \newcommand{\cplpagenumber}{xxxxxx}
 \newcommand{\cplpage}{\cplpagenumber-\thepage}

\begin{document}

\begin{CJK*}{UTF8}{gbsn}\vspace* {-4mm} \begin{center}
\large\bf{\boldmath{$\mathcal{O}_k$ null test with multi-task Gaussian processes: cosmic curvature and data compatibility}}
\footnotetext{\hspace*{-5.4mm}$^{*}$Corresponding authors. Email: gongyungui@nbu.edu.cn; yz@bnu.edu.cn

\noindent\copyright\,{\cplyear}
\href{http://www.cps-net.org.cn}{Chinese Physical Society} and
\href{http://www.iop.org}{IOP Publishing Ltd}
}
\\[5mm]
\normalsize \rm{}Yungui Gong (龚云贵)$^{1*}$\orcidlink{0000-0001-5065-2259}, Qing Gao (郜青)$^{2}$\orcidlink{0000-0003-3797-4370}, Xuchen Lu (路旭晨)$^{1}$\orcidlink{0000-0002-9093-9059}, Zhu Yi (易竹)$^{3*}$\orcidlink{0000-0001-7770-9542}
\\[3mm]\small\sl $^{1}$Institute of Fundamental Physics and Quantum Technology, Department of Physics,\\ School of Physical Science and Technology, Ningbo University, Ningbo, Zhejiang 315211, China

$^{2}$School of Physical Science and Technology, Southwest University, Chongqing 400715, China

$^{3}$Faculty of Arts and Sciences, Beijing Normal University, Zhuhai 519087, China
\\[4mm]\normalsize\rm{}(Received xxx; accepted manuscript online xxx)
\end{center}
\end{CJK*}

\vskip 1.5mm

\small{\narrower 
The $\mathcal{O}_k$ null test can not only assess whether the cosmic curvature is zero, therefore if true reducing  degeneracies between cosmic curvature and other cosmological parameters,  
but also provide a model-independent check of compatibility between different data sets. 
However, traditional implementations often require absolute distance data from Type Ia supernovae (SNe Ia) or baryon acoustic oscillation (BAO) measurements, 
limiting their applicability because such absolute distance data usually are not accessible. 
The BAO Alcock Paczynski (AP) parameter $F_{AP}$  is a measurement of a distance ratio, 
making the Dark Energy Spectroscopic Instrument (DESI) AP measurements particularly well suited for the  $\mathcal{O}_k$ null test  because no absolute distance measurements are required. 
We propose a novel null test of cosmic curvature tailored to DESI BAO data that combines $F_{AP}$ with ratios such as   $D_V'/D_V$ or $D_M'/D_M$. 
Crucially, this construction eliminates the need for absolute distance measurements.
We further develop multi-task Gaussian processes to perform the null test.  
This approach can also be applied to a joint DESI BAO and SNe Ia dataset, 
and we find that DESI BAO and SNe Ia data are compatible.
Although there is $\sim 2\sigma$ evidence of nonzero curvature at low redshift $z\lesssim 0.5$, 
 this result is not conclusive largely  due to the lack of observational data in the corresponding redshift range.

\par}\vskip 3mm
\normalsize\noindent{\narrower{DOI: \href{http://dx.doi.org/10.1088/0256-307X/43/5/051101}{10.1088/0256-307X/43/5/051101}}

\par}\vskip 5mm

\begin{multicols}{2}


Since the discovery of the accelerated expansion of the Universe in 1998 from observations of Type Ia supernovae (SNe Ia),\ucite{SupernovaSearchTeam:1998fmf, SupernovaCosmologyProject:1998vns} 
the concept of dark energy has been introduced as a phenomenological explanation for cosmic acceleration. 
The cosmological constant $\Lambda$ is the simplest candidate and fits naturally into Einstein's field equations. 
However, theoretical estimates of $\Lambda$ based on vacuum energy density exceed the observational value by many orders of magnitude,
a severe discrepancy known as the cosmological constant problem.\ucite{Weinberg:1988cp} Despite we know that dark energy has negative pressure and a gravitational interaction that behaves as repulsive rather than attractive, 
the nature of dark energy remains unknown. 
A central question in contemporary cosmology is whether dark energy is a true cosmological constant or a dynamical component whose equation of state evolves with time.

Observationally, efforts to distinguish a constant $\Lambda$ from dynamical dark energy commonly use parameterizations of the dark energy equation of state $w=p/\rho$. 
A widely used form is the Chevallier-Polarski-Linder (CPL) parametrization,\ucite{Chevallier:2000qy,Linder:2002et} $w(a)=w_0+w_a(1-a)$  with $a$ being the scale factor, 
which captures a broad class of scalar-field dynamics 
and allows for a time-varying equation of state. 
Recent joint analyses combining baryon acoustic oscillation (BAO) measurements from the Dark Energy Spectroscopic Instrument (DESI), \ucite{DESI:2024mwx,DESI:2025zgx} 
cosmic microwave background (CMB) data from Planck,\ucite{Planck:2018vyg} and SNe Ia compilations such as Pantheon Plus,\ucite{Scolnic:2021amr} Union3,\ucite{Rubin:2023jdq} 
and the five-year Dark Energy Survey (DES Y5) results \ucite{DES:2024jxu,DES:2025sig} have reported evidence favoring dynamical dark energy over a pure cosmological constant. 
If confirmed, these results would have profound implications for fundamental physics and our understanding of cosmic evolution.

The evidence for cosmic acceleration and dynamical dark energy, 
however, depends sensitively on models and combinations of data. 
To make the evidence more reliable and robust, this paper addresses two important questions: (1) can we probe the Universe in a model-independent way? 
and (2) are different data sets mutually compatible so that they can be combined to infer cosmological evolution and matter contents reliably?
There is an extensive literature on model-independent reconstructions of cosmic expansion.\ucite{Visser:1997qk,Visser:1997tq,Gong:2007fm,Gong:2007zf,Santos:2006ja,Santos:2007pp,Yang:2019fjt,Lu:2024hvv,Gao:2025ozb,Holsclaw:2010nb,Holsclaw:2010sk,Holsclaw:2011wi,Bilicki:2012ub,Seikel:2012uu,Seikel:2012cs,Clarkson:2007pz,Zunckel:2008ti,Cai:2015pia,Gong:2006tx,Gong:2006gs,Nesseris:2010ep,Clarkson:2010bm,Shafieloo:2012ht,Shafieloo:2012rs,Gao:2012ef,Gong:2013bn,Yahya:2013xma,Sahni:2014ooa,Li:2015nta,Vitenti:2015aaa,Zhang:2016tto,Wei:2016xti,Yu:2017iju,Yennapureddy:2017vvb,Velten:2017ire,Marra:2017pst,Melia:2018tzi,Gomez-Valent:2018hwc,Haridasu:2018gqm,Capozziello:2018jya,Arjona:2019fwb,Jesus:2019nnk,Franco:2019wbj,Dhawan:2021mel,Gangopadhyay:2023nli,Sharma:2024mtq,Dinda:2024kjf,Jiang:2024xnu,Ghosh:2024kyd,Cortes:2024lgw,Shlivko:2024llw,deCruzPerez:2024shj,Roy:2024kni,Chatrchyan:2024xjj,Perivolaropoulos:2024yxv,Payeur:2024dnq,Chan-GyungPark:2024brx,Gao:2024ily,Dinda:2024ktd}
For example, from the definition of cosmic deceleration, $q=-\ddot{a}/(aH)>0$, 
one obtains the constraints on the Hubble parameter $H(z)=\dot{a}/a\ge H_0(1+z)$ or the luminosity distance $H_0 d_L\le (1+z)\ln(1+z)$ for a spatially flat universe with $\Omega_k=0$, 
here an overdot means the derivative with respect to the cosmic time $t$, a subscript $0$ means the value of the quantity at present, the redshift $z=a_0/a-1$  and $H_0=100 h$ km/s Mpc$^{-1}$ is the Hubble constant. 
Thus, the measurements on the luminosity distance $d_L$ or  the dimensionless Hubble parameter $E(z)=H(z)/H_0$ from SNe Ia can provide direct, 
model-independent evidence of cosmic acceleration without assuming a particular cosmological model or even a specific theory of gravity.\ucite{Visser:1997qk,Visser:1997tq,Gong:2007fm,Gong:2007zf,Santos:2006ja,Santos:2007pp,Yang:2019fjt}
 This is the null test of cosmic acceleration. 
In practice, however, the zero-point calibration problem of SNe Ia limits the direct application of such model-independent  null tests.

Fortunately, DESI BAO data \ucite{DESI:2025zgx}  provide measurements on the ratio between the transverse comoving distance to the sound horizon at the drag epoch $D_M(z)/r_d$,  the ratio between the line-of-sight distance to the sound horizon at the drag epoch  $D_H(z)/r_d$, and their correlations;
or the Alcock-Paczynski (AP) parameter  $F_{AP}=D_M/D_H$, the ratio between the isotropic BAO distance to the sound horizon at the drag epoch $D_V/r_d$, and their correlations;
so one can use the AP parameter to reconstruct $E(z)$ for a spatially flat
universe \ucite{Lu:2024hvv,Gao:2025ozb} via Gaussian Process (GP) regression \ucite{Holsclaw:2010nb,Holsclaw:2010sk,Holsclaw:2011wi,Bilicki:2012ub,Seikel:2012uu,Seikel:2012cs}  to perform a null test of cosmic acceleration. 
The line-of-sight distance variable is defined as
\begin{equation}
\label{eq:D_H}
    D_H=\frac{c}{H(z)},
\end{equation}
where $c$ is the speed of light.
The sound horizon at the drag epoch $z_d$ is
\begin{equation}
\label{eq:r_d}
r_d=\int_{z_d}^\infty \frac{c_s(z)}{H(z)}d z,
\end{equation}
where $c_s(z)$ is the speed of sound in the baryon-photon plasma.
The transverse comoving distance $D_{M}$ is given by
\begin{equation}
\label{eq:D_M}
    D_{M}(z)=\frac{c}{H_0 \sqrt{|\Omega_{k0}|}} \text{sinn} \left[\sqrt{|\Omega_{k0}|} \int_0^z \frac{d z^{\prime}}{E\left(z^{\prime}\right)}\right],
\end{equation}
where $\text{sinn}(\sqrt{|\Omega_{k0}|}x)/\sqrt{|\Omega_{k0}|}=\sinh(\sqrt{|\Omega_{k0}|}x)/\sqrt{|\Omega_{k0}|}$, $x$,
and $\sin(\sqrt{|\Omega_{k0}|}x)/\sqrt{|\Omega_{k0}|}$ for $\Omega_{k0}>0$, $\Omega_{k0}=0$, and $\Omega_{k0}<0$,
respectively. 
The isotropic BAO distance is  defined as
\begin{equation}
    D_V(z)=[zD_M^2(z)D_H(z)]^{1/3}.
\end{equation}
Note that at $z=0$,  $D_M(z)=F_{AP}(z)=D_V(z)=0$, and from Eq. \eqref{eq:D_H} we can use the data $D_H/r_d$ to get
\begin{equation}
\label{eq:h0rd}  
H_0 r_d=\frac{c}{D_H(z=0)/r_d}.
\end{equation}
Because $F_{AP}$ is a ratio which is independend of the (unknown) sound horizon $r_d$,
and therefore avoids the SNe Ia zero-point issue when providing model-independent evidence for cosmic acceleration.\ucite{Lu:2024hvv,Gao:2025ozb}

With nonzero cosmic curvature $\Omega_k$, the phantom crossing favored by DESI BAO data can be avoided,\ucite{Dinda:2025iaq,Chen:2025mlf}
so determining whether $\Omega_k=0$ is also crucial.
The $\mathcal{O}_k$ diagnostic,\ucite{Clarkson:2007pz, Zunckel:2008ti,Cai:2015pia, Gao:2025ozb} $\mathcal{O}_k(z)=E(z)D'(z)-1$,\ucite{Cai:2015pia} offers a model-independent test of spatial flatness, where $D'(z)=dD(z)/dz$
and $D(z)=H_0D_M(z)$. 
Applying this diagnostic typically uses SNe Ia to reconstruct $D(z)$, 
which reintroduces the zero-point calibration problem.
To avoid that and exploit DESI BAO AP data, 
the $\mathcal{O}_k$ diagnostic can be recast as $\mathcal{O}_k(z)=F_{AP}(z)D'_M(z)/D_M(z)-1$.\ucite{Gao:2025ozb}
In this expression, the dependence on the sound horizon $r_d$ cancels in the ratios  $D'_M/D_M$  and $D_M/D_H$ constructed from DESI BAO data, 
so the null test of cosmic curvature depends only on the assumption of the cosmological principle.
If $\mathcal{O}_k(z)=0$, then the model-independent implications are: 
1) the cosmic curvature is zero, removing concerns about the degeneracy between $\Omega_k$ and $w$;
2) the data sets used to reconstruct $\mathcal{O}_k(z)=0$ are compatible.
Moreover, one can reconstruct the ratio $D'_M/D_M$ from SNe Ia without suffering the zero-point calibration problem. 
Combining DESI BAO AP data with the ratio $D'_M/D_M$ derived from SNe Ia,  therefore not only provides a null test of the cosmic curvature, 
but also a model-independent test of the  compatibility between DESI BAO and SNe Ia data. 

Because DESI Data Release 2 (DR2) BAO measurements are reported either as $(D_M/r_d, D_H/r_d)$ with correlations or as $(F_{AP}, D_V/r_d)$ with correlations,
the simple combination of $F_{AP}$ and $D'_M/D_M$,   the reconstruction of $F_{AP}$ from DESI BAO AP data
and the reconstruction of $D_M/r_d$ and $D_M'/r_d$ from the DESI BAO data $D_M/r_d$ by single-task GP (SGP)  do not properly account for these correlations. 
A more consistent approach is to reconstruct $F_{AP}$
from $F_{AP}$ and $D_V/r_d$ joint data including their correlations using multi-task GP (MGP) method developed in the appendix, 
and to use $F_{AP}$ together with $D'_V/D_V$ in the $\mathcal{O}_k$ diagnostic so correlations are handled correctly.
In this work we do this and use DESI DR2 BAO AP and SNe Ia data to perform a null test of spatial flatness and test the consistency between data.

We first use the MGP method to check the consistency in DESI BAO data by reconstructing $F_{AP}$ from AP data only, 
from joint data $F_{AP}$, $D_V/r_d$ and their correlations, and from  joint data   $D_M/r_d$, $D_H/r_d$ and their correlations. 
 The details of the MGP reconstruction by including the correlations are presented in the appendix.  
The reconstructed results are shown in Fig. \fref{fig:fap}{1}.
As we can see from Fig. \fref{fig:fap}{1}, 
the results from these three methods are consistent with each other even when the correlations are not considered,
so the DESI BAO DR2 data is self-consistent.

For a spatially flat universe with $\Omega_{k0}=0$,
\begin{equation}
\label{eq:D_M_flat}
D_M(z)=\frac{c}{H_0}\int_{0}^{z}\frac{d z^{\prime}}{E\left(z^{\prime}\right)}dz^{\prime},
\end{equation}
so $D_M'(z)=D_H(z)$ and $F'_{AP}(z=0)=1$.
Taking this relation into account and considering the correlation between $D_M$ and $D_H$ in the DESI BAO DR2 dataset,
we reconstruct $D_M$ and $D_H=D_M'$ with MGP, and the result is shown as the green dashed line in Fig. \fref{fig:gpdmdh}{2}.
Substituting the MGP result from Fig. \fref{fig:gpdmdh}{2} into Eq. \eqref{eq:h0rd}, we get $r_d h= 100.61\pm 1.62$ Mpc under the assumption of $\Omega_k=0$.
This result is consistent with Planck 2018 value $r_d h= 98.82\pm 0.82$ Mpc which is based on flat $\Lambda$CDM model,\ucite{Planck:2018vyg} 
so Placnk data is consistent with DESI BAO data if $\Omega_k=0$. 

\vskip 4mm

\fl{fig:fap}\centerline{\includegraphics[width=0.9\linewidth]{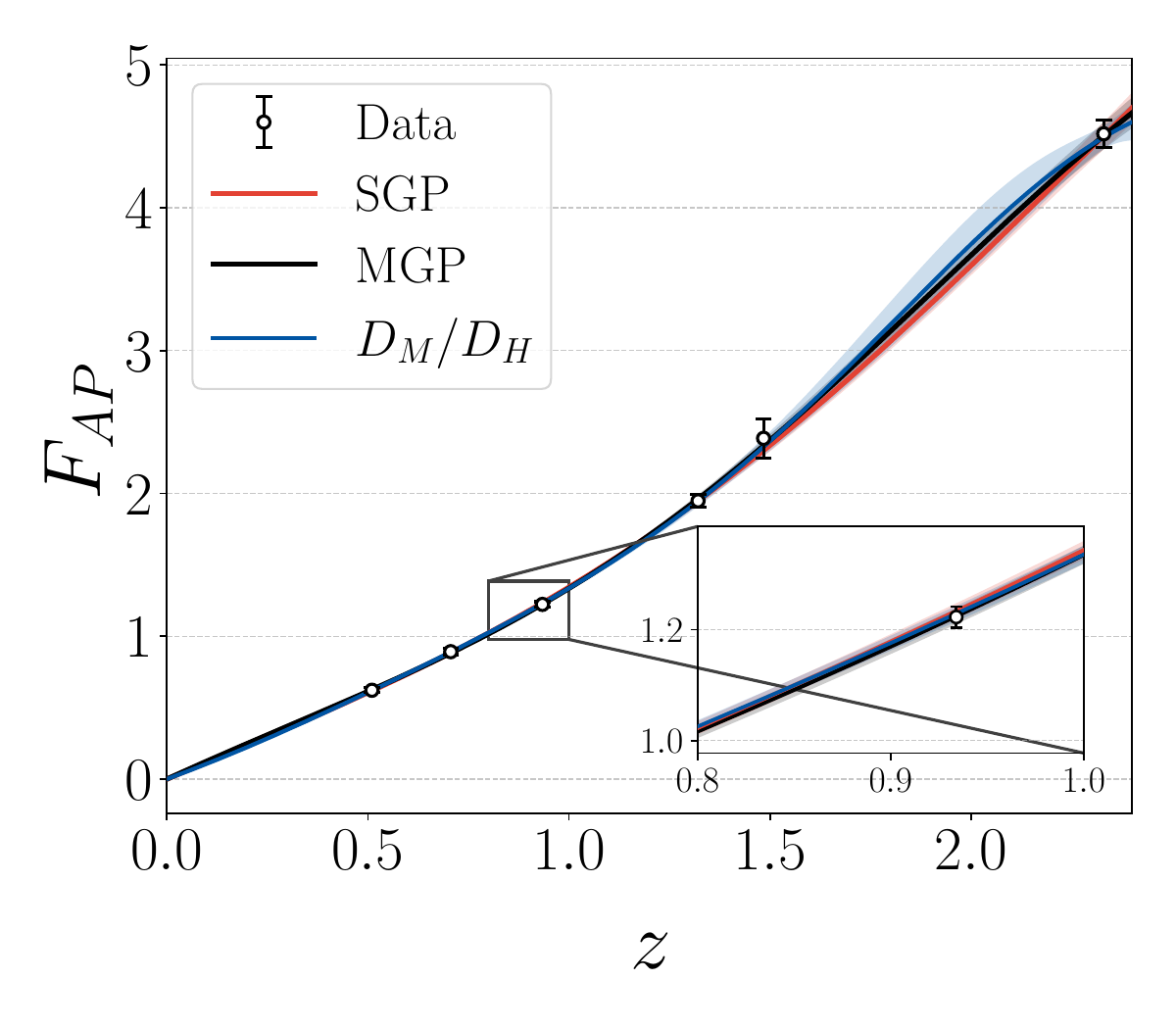}}
\vskip 2mm

\figcaption{7.5}{1}{
The reconstructed $F_{AP}$ along with the $1\sigma$ error from DESI BAO DR2 data.
The constraints $D_M(z=0)=0$ and $F_{AP}(z=0)=0$ are imposed.
The red line labeled as SGP is reconstructed from the data $F_{AP}$ only.
The black line labeled as MGP is reconstructed from the joint data $F_{AP}$, $D_V/r_d$ and their correlations.
The blue line labeled as $D_M/D_H$ is reconstructed from the joint data $D_M/r_d$, $D_H/r_d$ and their correlations.
}
\medskip

\vskip 4mm

\fl{fig:gpdmdh}\centerline{\includegraphics[width=0.9\linewidth]{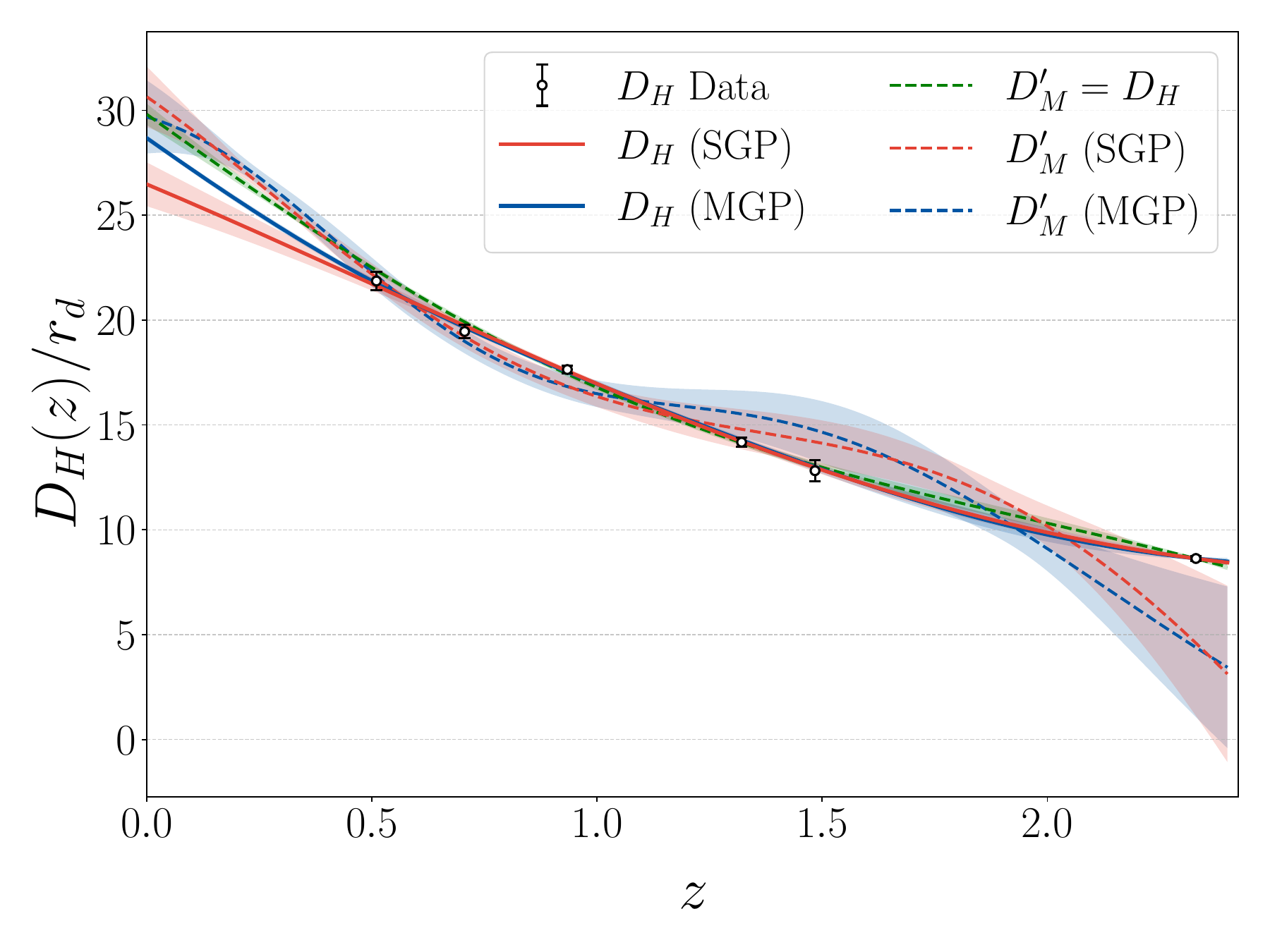}}
\vskip 2mm

\figcaption{7.5}{2}{
The reconstructed $D_H/r_d$ and $D_M'/r_d$ along with the $1\sigma$ error from DESI BAO DR2 data. 
The constraint $D_M(z=0)=0$ is imposed.
The green line labeled as $D_M'=D_H$ is reconstructed with the data $D_M/r_d$, $D_H/r_d$ and their correlations by taking the data $D_H/r_d$ as $D_M'/r_d$.
The blue lines labeled as MGP are reconstructed with the data $D_M$, $D_H$ and their correlations.
The red lines are reconstructed separately from the respective data alone.
The solid lines are for $D_H/r_d$ and the dashed lines are for $D_M'/r_d$.}

\medskip

Now we discuss the null test of spatial curvature.
Based on the relation $D_M'=D_H$ for a spatially flat universe and the independence of $r_d$ from $F_{AP}$, the $\mathcal{O}_k$ null test was proposed,\ucite{Gao:2025ozb}
\begin{equation}
\label{oktest1}
\mathcal{O}_k(z)=F_{AP}(z)\frac{D_M'(z)}{D_M(z)}-1.
\end{equation}
The ratio $D_M'/D_M$ cleverly eliminates the dependence on the sound horizon.
However, DESI DR2 BAO data either measures $D_M/r_d$ and $D_H/r_d$,
or $F_{AP}$ and $D_V/r_d$, i.e., either $D_M$ and $D_H$ are correlated or $F_{AP}$ and $D_V$ are correlated, so the combination of $F_{AP}$ and $D_M$ does not properly account the correlation between $F_{AP}$ and $D_M$.
Instead we propose the $\mathcal{O}_k$ null test
\begin{equation}
\label{eq:fap_okz}
\mathcal{O}_k=F_{AP}\frac{D'_V}{D_V}+\frac{1}{3}F'_{AP}-\frac{1}{3}\frac{F_{AP}}{z}-1
=0.
\end{equation}
The main difference between this work and our previous work (Ref. \cite{Gao:2025ozb}) lies in the development of the MGP processes and the simultaneous reconstruction of the functions  $F_{AP} $, $D_V/r_d$ and their derivatives or $D_M/r_d$, $D_H/r_d$ and their derivatives from the joint data sets $F_{AP}$, $D_V/r_d$ and their covariance, or $D_M/r_d$, $D_H/r_d$ and their covariance, which are used to perform the null test.

\vskip 4mm

\fl{fig:okz_fapdv}\centerline{\includegraphics[width=0.9\linewidth]{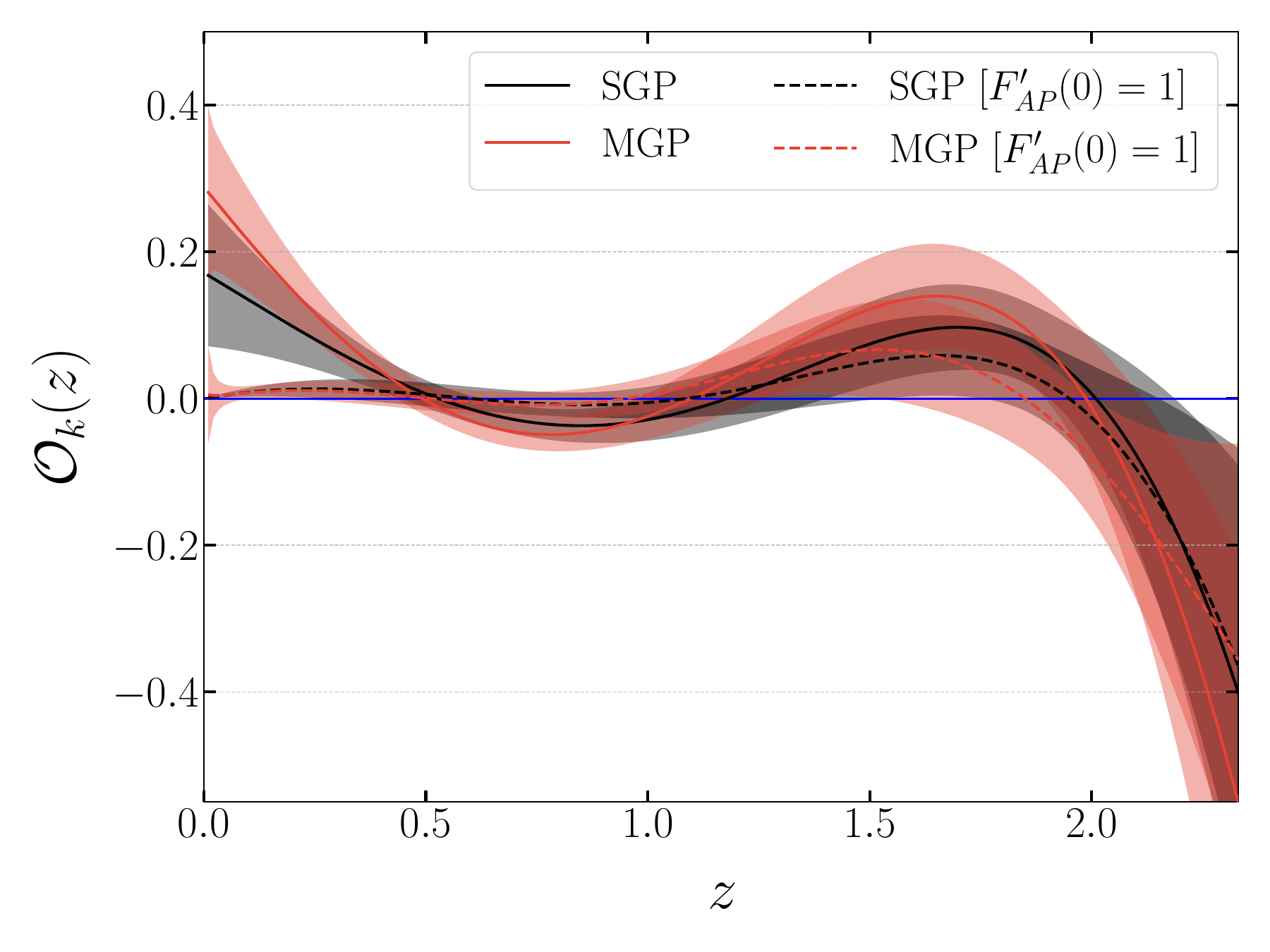}}
\vskip 2mm

\figcaption{7.5}{3}{
The reconstructed $\mathcal{O}_k$ with its $1\sigma$ uncertainty from DESI BAO measurements of $F_{\rm AP}$ and $D_V/r_d$.
The constraints $D_V(z=0)=0$ and $F_{AP}(z=0)=0$ are imposed.
Solid curves show the posterior mean; shaded regions indicate the $1\sigma$ uncertainties.
We also impose the constraint $F'_{AP}(z=0)=1$ and the results are shown with dashed curves.
SGP corresponds to reconstructing $F_{\rm AP}$ and $D_V/r_d$ separately from the respective data sets $F_{\rm AP}$ and $D_V/r_d$.
MGP denotes the reconstruction with joint data  $F_{\rm AP}$, $D_V/r_d$ and their correlations.}
\medskip

Combining DESI BAO DR2 data $F_{\rm AP}$, $D_V/r_d$, and their correlations, 
we calculate $\mathcal{O}_k$ using Eq. \eqref{eq:fap_okz}. 
The results are shown in Fig. \fref{fig:okz_fapdv}{3}. 
For comparison, we also show the result obtained when the correlations are ignored, i.e., $F_{\rm AP}$ and $D_V/r_d$ are reconstructed separately from their respective data without considering their correlations.
From  Fig. \fref{fig:okz_fapdv}{3}, we see that $\Omega_k\neq 0$ at $z\lesssim 0.5$.
The reconstructed results with and without considering the correlations are similar.
Due to the degeneracy between $\Omega_k$ and $w$, 
the nonzero spatial curvature at $z\lesssim 0.5$ and  
the phantom crossing of dark energy may have the same origin,
suggesting that the phantom crossing preferred by DESI BAO data could be an artifact of the flat CPL parametrization.
However, because there is no BAO data in the redshift region $z\lesssim 0.5$,
the reconstructed results may not reliable.
If we impose the constraint $F'_{AP}(0)=1$, we find that the results as shown in Fig. \fref{fig:okz_fapdv}{3} are consistent with zero curvature. 
Therefore, we suspect that the apparent nonzero curvature at $z\lesssim 0.5$ may be caused by the lack of data in DESI DR2 in that redshift range.

To test whether the deviation from the spatial flatness at $z\lesssim 0.5$ is caused by the lack of data in that redshift regions, we check whether $D_M'=D_H$ in the data.
We first reconstruct $D_M'/r_d$ from the $D_M/r_d$ data,  and $D_H/r_d$ from the $D_H/r_d$.
The results are shown in Fig. \fref{fig:gpdmdh}{2} with red lines. 
Then we reconstruct $D_M'/r_d$ and $D_H/r_d$ from the joint data $D_M/r_d$, $D_H/r_d$ and their correlations to check whether $D_H=D_M'$.
The results, labeled as MGP, are shown in Fig. \fref{fig:gpdmdh}{2}.
From Fig. \fref{fig:gpdmdh}{2}, 
we see that $D_H(z)=D_M'(z)$ at $z\gtrsim 0.5$ even though we don't consider the correlations;
but $D_H\neq D_M'$ at $z\lesssim 0.5$ even though we reconstruct them jointly and consider the correlations between $D_M$ and $D_H$.
The reconstructed results with and without including the correlations are similar.
 Therefore, at $z\lesssim 0.5$, $\mathcal{O}_k\neq 0$ may due to $D_H\neq D_M'$ in the DESI BAO data and is caused by the lack of data in this redshift region. 

\vskip 5mm

\fl{fig:okz_fapsn}\centerline{\includegraphics[width=0.9\linewidth]{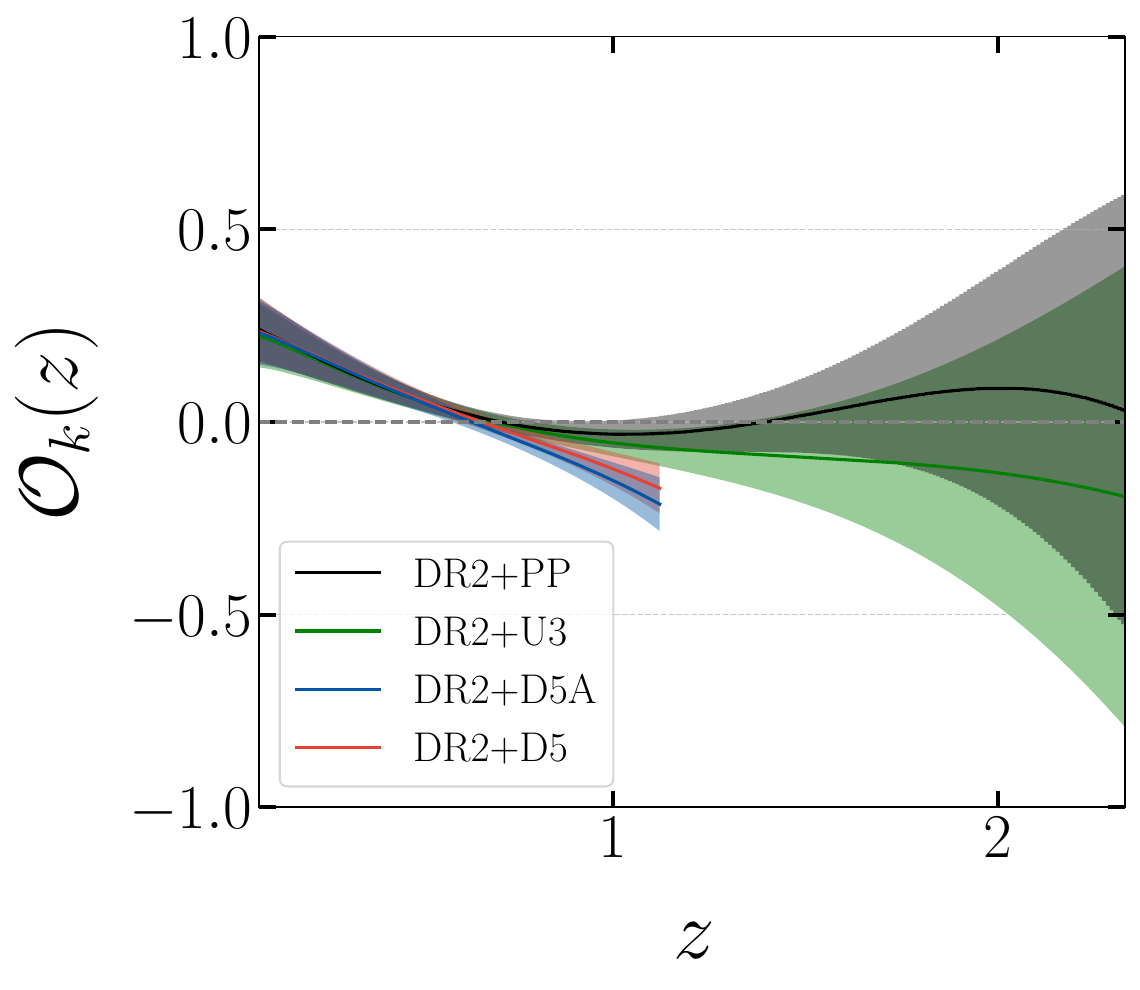}}
\vskip 2mm

\figcaption{7.5}{4}{
The reconstructed $\mathcal{O}_k$ along with the $1\sigma$ errors from DESI BAO AP and SNe Ia.
The constraints $F_{AP}(z=0)=0$ is imposed. D5A and D5 are for the original and recalibrated DES Y5 data, respectively. }
\medskip

For the null test \eqref{oktest1}, we can also combine DESI BAO and SNe Ia data. 
We use three SNe Ia samples: 
the Pantheon Plus sample of 1550 spectroscopically confirmed SNe Ia,\ucite{Scolnic:2021amr}
the Union3 compilation of 2087 SNe Ia,\ucite{Rubin:2023jdq} 
and the Dark Energy Survey (DES) five-year compilation of 1829 SNe Ia.\ucite{DES:2024jxu,DES:2025sig} 
For brevity we denote these datasets PP (Pantheon Plus), U3 (Union3), and D5 (DES Y5), respectively.
Combining DESI BAO DR2 $F_{AP}$ data with each SNe Ia sample, 
we compute $\mathcal{O}_k$ using Eq. \eqref{oktest1} and the results are shown in Fig. \fref{fig:okz_fapsn}{4}. 
Fig. \fref{fig:okz_fapsn}{4} again shows $\Omega_k \neq 0$ at $z \lesssim 0.5$,
suggesting that the deviation is driven by the  lack of BAO data at low redshift region. 
Additional observations, especially more low-$z$ BAO measurements,
are therefore needed to robustly test the spatial flatness.
Unlike U3 and PP, the original D5 sample (D5A) also indicates a deviation from $\Omega_k=0$ at $z\gtrsim 0.5$,\ucite{DES:2024jxu} 
but with recalibrated D5 sample,\ucite{DES:2025sig} the deviation becomes smaller.
Therefore, the $\mathcal{O}_k$ null test is not only robust to check the spatial flatness, but also powerful to detect problems in data.

Finally, we further examine the consistency between BAO and SNe Ia datasets from the ratio $D_M/D_M'$.
We reconstruct $D_M/D_M'$ using the three SNe Ia samples: U3, PP, and D5, using the joint BAO measurements $D_M/r_d$, $D_H/r_d$ and their correlations, 
and using $D_M/r_d$ alone. 
The resulting comparisons of $D_M/D_M'$ are shown in Fig. \fref{fig:datacomp}{5}. 
From Fig. \fref{fig:datacomp}{5}, we see that the SNe Ia and BAO data are compatible  with each other, and it confirms the results obtained from the $\mathcal{O}_k$ null test.

\vskip 4mm

\fl{fig:datacomp}\centerline{\includegraphics[width=0.9\linewidth]{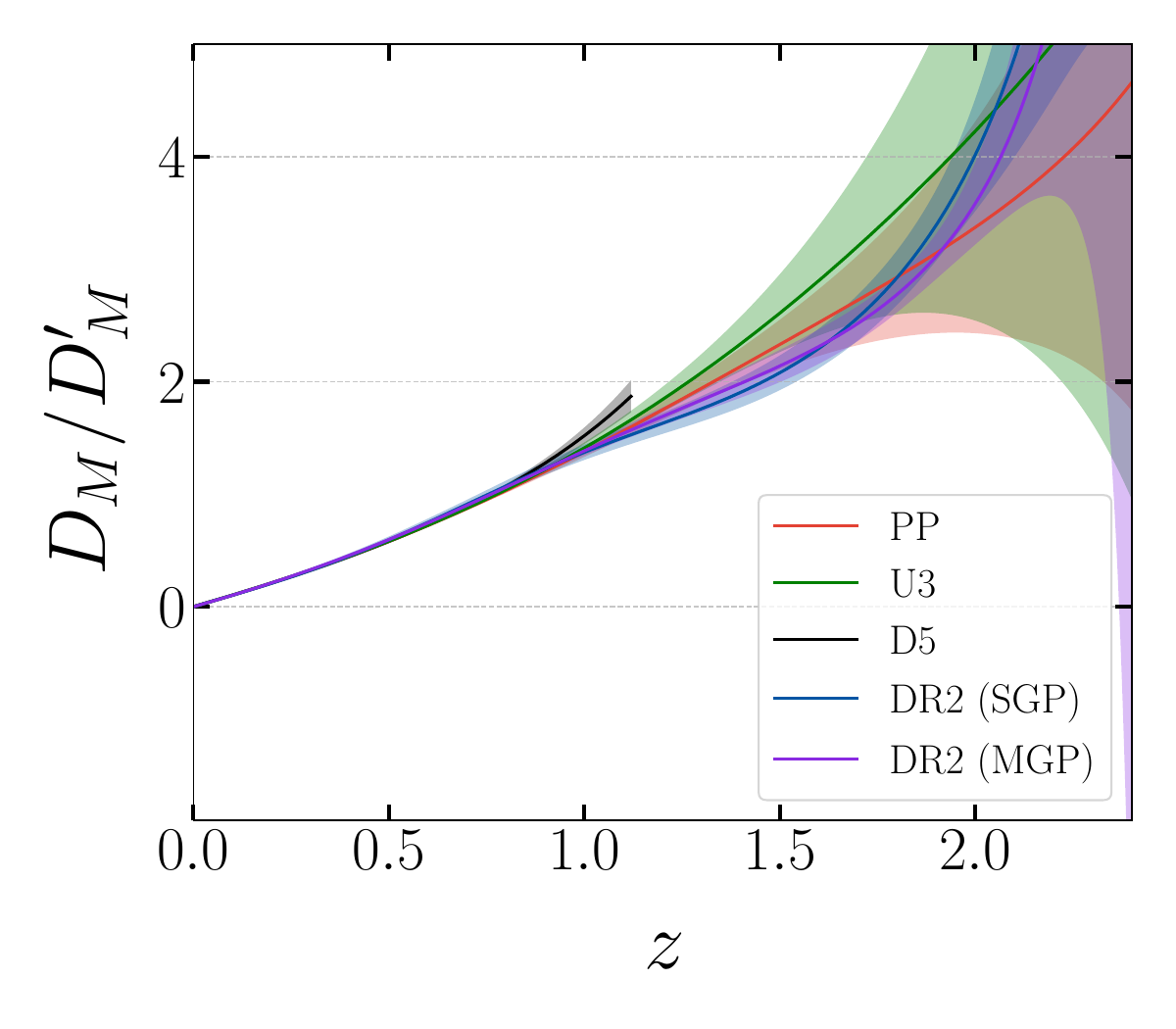}}
\vskip 2mm

\figcaption{7.5}{5}{
The reconstructed $D_M/D_M'$ along with the $1\sigma$ errors from DESI BAO and SNe Ia.
The constraints $D_M(z=0)=0$ is imposed for the reconstruction of DESI BAO data.
The blue line labeled as SGP is reconstructed from the data $D_M$ only,
the purple line labeled as MGP is reconstructed from the joint data $D_M$, $D_H$ and their correlations.}
\medskip

In conclusion, we propose novel null tests of cosmic curvature specifically tailored to DESI BAO data:
$\mathcal{O}_k(z)=F_{\rm AP}(z)\,D_M'(z)/D_M(z)-1$ and
$\mathcal{O}_k=F_{\rm AP}\,D'_V/D_V+F'_{\rm AP}/3-F_{\rm AP}/(3z)-1$.
Our reconstructions with MGP indicate that a spatially flat universe is disfavored by DESI DR2 BAO data at low redshift $z \lesssim 0.5$, a region where BAO measurements are currently sparse. 
This pattern suggests the apparent deviation from flatness,
and even the phantom crossing of dark energy, may be driven by incomplete low-$z$ BAO coverage rather than a genuine physical consequence. 
Therefore, additional observations, particularly more BAO measurements at low redshift, 
are required to test spatial flatness robustly.
Reconstructions performed with and without including the full correlations yield similar results. 
We also find that the DESI BAO DR2 data are self-consistent, 
and that the SNe Ia datasets are broadly consistent with DESI DR2 BAO data. 
The $\mathcal{O}_k$ null test is not only robust for testing spatial flatness, 
but also a powerful tool for assessing data compatibility and uncovering potential data issues.

\textit{Acknowledgements.} This research is supported in part by 
the National Natural Science Foundation of China Grant Nos. 12588101, 12535002,  12175184, and 12205015.

{\it Notes Added}: While we were developing a GP method to jointly reconstruct two functions from two data sets including their correlations, the paper \ucite{Dinda:2025hiu} appeared. 
The idea and preliminary results for reconstructing $F_{AP}$ without accounting for correlations to test the compatibility between DESI BAO data and SNe Ia data, 
were presented in talks by one of the authors (YG) at Lanzhou University on September 19 and the University of Science and Technology of China on September 25.


\begin{thebibliography}{10}

\bibitem{SupernovaSearchTeam:1998fmf}
Riess A~G et al. (Supernova Search Team)
  \href{https://doi.org/10.1086/300499}{1998 {\em Astron. J.} {\bf 116} 1009}

\bibitem{SupernovaCosmologyProject:1998vns}
Perlmutter S et al. (Supernova Cosmology Project)
  \href{https://doi.org/10.1086/307221}{1999 {\em Astrophys. J.} {\bf 517} 565}

\bibitem{Weinberg:1988cp}
Weinberg S \href{https://doi.org/10.1103/RevModPhys.61.1}{1989 {\em Rev. Mod.
  Phys.} {\bf 61} 1}

\bibitem{Chevallier:2000qy}
Chevallier M and Polarski D
  \href{https://doi.org/10.1142/S0218271801000822}{2001 {\em Int. J. Mod. Phys.
  D} {\bf 10} 213}

\bibitem{Linder:2002et}
Linder E~V \href{https://doi.org/10.1103/PhysRevLett.90.091301}{2003 {\em Phys.
  Rev. Lett.} {\bf 90} 091301}

\bibitem{DESI:2024mwx}
Adame A~G et al. (DESI)
  \href{https://doi.org/10.1088/1475-7516/2025/02/021}{2025 {\em JCAP} {\bf 02}
  021}

\bibitem{DESI:2025zgx}
Abdul~Karim M et al. (DESI) \href{https://doi.org/10.1103/tr6y-kpc6}{2025 {\em
  Phys. Rev. D} {\bf 112} 083515}

\bibitem{Planck:2018vyg}
Aghanim N et al. (Planck)
  \href{https://doi.org/10.1051/0004-6361/201833910}{2020 {\em Astron.
  Astrophys.} {\bf 641} A6} [Erratum: Astron.Astrophys. 652, C4 (2021)]

\bibitem{Scolnic:2021amr}
Scolnic D et al. \href{https://doi.org/10.3847/1538-4357/ac8b7a}{2022 {\em
  Astrophys. J.} {\bf 938} 113}

\bibitem{Rubin:2023jdq}
Rubin D et al. \href{https://doi.org/10.3847/1538-4357/adc0a5}{2025 {\em
  Astrophys. J.} {\bf 986} 231}

\bibitem{DES:2024jxu}
Abbott T M~C et al. (DES) \href{https://doi.org/10.3847/2041-8213/ad6f9f}{2024
  {\em Astrophys. J. Lett.} {\bf 973} L14}

\bibitem{DES:2025sig}
Popovic B et al. (DES) \href{https://arxiv.org/abs/2511.07517}{2025
  arXiv:2511.07517}

\bibitem{Visser:1997qk}
Visser M \href{https://doi.org/10.1126/science.276.5309.88}{1997 {\em Science}
  {\bf 276} 88}

\bibitem{Visser:1997tq}
Visser M \href{https://doi.org/10.1103/PhysRevD.56.7578}{1997 {\em Phys. Rev.
  D} {\bf 56} 7578}

\bibitem{Gong:2007fm}
Gong Y and Wang A \href{https://doi.org/10.1016/j.physletb.2007.06.065}{2007
  {\em Phys. Lett. B} {\bf 652} 63}

\bibitem{Gong:2007zf}
Gong Y, Wang A, Wu Q, and Zhang Y-Z
  \href{https://doi.org/10.1088/1475-7516/2007/08/018}{2007 {\em JCAP} {\bf 08}
  018}

\bibitem{Santos:2006ja}
Santos J, Alcaniz J~S, and Reboucas M~J
  \href{https://doi.org/10.1103/PhysRevD.74.067301}{2006 {\em Phys. Rev. D}
  {\bf 74} 067301}

\bibitem{Santos:2007pp}
Santos J, Alcaniz J~S, Pires N, and Reboucas M~J
  \href{https://doi.org/10.1103/PhysRevD.75.083523}{2007 {\em Phys. Rev. D}
  {\bf 75} 083523}

\bibitem{Yang:2019fjt}
Yang Y and Gong Y \href{https://doi.org/10.1088/1475-7516/2020/06/059}{2020
  {\em JCAP} {\bf 06} 059}

\bibitem{Lu:2024hvv}
Lu X, Gao S, and Gong Y
  \href{https://doi.org/10.1088/0256-307X/43/1/011101}{2026 {\em Chin. Phys.
  Lett.} {\bf 43} 011101}

\bibitem{Gao:2025ozb}
Gao S, Gao Q, Gong Y, and Lu X
  \href{https://doi.org/10.1007/s11433-025-2682-1}{2025 {\em Sci. China Phys.
  Mech. Astron.} {\bf 68} 280408}

\bibitem{Holsclaw:2010nb}
Holsclaw T, Alam U, Sanso B, Lee H, Heitmann K, Habib S, and Higdon D
  \href{https://doi.org/10.1103/PhysRevD.82.103502}{2010 {\em Phys. Rev. D}
  {\bf 82} 103502}

\bibitem{Holsclaw:2010sk}
Holsclaw T, Alam U, Sanso B, Lee H, Heitmann K, Habib S, and Higdon D
  \href{https://doi.org/10.1103/PhysRevLett.105.241302}{2010 {\em Phys. Rev.
  Lett.} {\bf 105} 241302}

\bibitem{Holsclaw:2011wi}
Holsclaw T, Alam U, Sanso B, Lee H, Heitmann K, Habib S, and Higdon D
  \href{https://doi.org/10.1103/PhysRevD.84.083501}{2011 {\em Phys. Rev. D}
  {\bf 84} 083501}

\bibitem{Bilicki:2012ub}
Bilicki M and Seikel M
  \href{https://doi.org/10.1111/j.1365-2966.2012.21575.x}{2012 {\em Mon. Not.
  Roy. Astron. Soc.} {\bf 425} 1664}

\bibitem{Seikel:2012uu}
Seikel M, Clarkson C, and Smith M
  \href{https://doi.org/10.1088/1475-7516/2012/06/036}{2012 {\em JCAP} {\bf 06}
  036}

\bibitem{Seikel:2012cs}
Seikel M, Yahya S, Maartens R, and Clarkson C
  \href{https://doi.org/10.1103/PhysRevD.86.083001}{2012 {\em Phys. Rev. D}
  {\bf 86} 083001}

\bibitem{Clarkson:2007pz}
Clarkson C, Bassett B, and Lu T H-C
  \href{https://doi.org/10.1103/PhysRevLett.101.011301}{2008 {\em Phys. Rev.
  Lett.} {\bf 101} 011301}

\bibitem{Zunckel:2008ti}
Zunckel C and Clarkson C
  \href{https://doi.org/10.1103/PhysRevLett.101.181301}{2008 {\em Phys. Rev.
  Lett.} {\bf 101} 181301}

\bibitem{Cai:2015pia}
Cai R-G, Guo Z-K, and Yang T
  \href{https://doi.org/10.1103/PhysRevD.93.043517}{2016 {\em Phys. Rev. D}
  {\bf 93} 043517}

\bibitem{Gong:2006tx}
Gong Y and Wang A \href{https://doi.org/10.1103/PhysRevD.73.083506}{2006 {\em
  Phys. Rev. D} {\bf 73} 083506}

\bibitem{Gong:2006gs}
Gong Y-G and Wang A \href{https://doi.org/10.1103/PhysRevD.75.043520}{2007 {\em
  Phys. Rev. D} {\bf 75} 043520}

\bibitem{Nesseris:2010ep}
Nesseris S and Shafieloo A
  \href{https://doi.org/10.1111/j.1365-2966.2010.17254.x}{2010 {\em Mon. Not.
  Roy. Astron. Soc.} {\bf 408} 1879}

\bibitem{Clarkson:2010bm}
Clarkson C and Zunckel C
  \href{https://doi.org/10.1103/PhysRevLett.104.211301}{2010 {\em Phys. Rev.
  Lett.} {\bf 104} 211301}

\bibitem{Shafieloo:2012ht}
Shafieloo A, Kim A~G, and Linder E~V
  \href{https://doi.org/10.1103/PhysRevD.85.123530}{2012 {\em Phys. Rev. D}
  {\bf 85} 123530}

\bibitem{Shafieloo:2012rs}
Shafieloo A, Sahni V, and Starobinsky A~A
  \href{https://doi.org/10.1103/PhysRevD.86.103527}{2012 {\em Phys. Rev. D}
  {\bf 86} 103527}

\bibitem{Gao:2012ef}
Gao Q and Gong Y \href{https://doi.org/10.1142/S0218271813500351}{2013 {\em
  Int. J. Mod. Phys. D} {\bf 22} 1350035}

\bibitem{Gong:2013bn}
Gong Y and Gao Q \href{https://doi.org/10.1140/epjc/s10052-014-2729-2}{2014
  {\em Eur. Phys. J. C} {\bf 74} 2729}

\bibitem{Yahya:2013xma}
Yahya S, Seikel M, Clarkson C, Maartens R, and Smith M
  \href{https://doi.org/10.1103/PhysRevD.89.023503}{2014 {\em Phys. Rev. D}
  {\bf 89} 023503}

\bibitem{Sahni:2014ooa}
Sahni V, Shafieloo A, and Starobinsky A~A
  \href{https://doi.org/10.1088/2041-8205/793/2/L40}{2014 {\em Astrophys. J.
  Lett.} {\bf 793} L40}

\bibitem{Li:2015nta}
Li Z, Gonzalez J~E, Yu H, Zhu Z-H, and Alcaniz J~S
  \href{https://doi.org/10.1103/PhysRevD.93.043014}{2016 {\em Phys. Rev. D}
  {\bf 93} 043014}

\bibitem{Vitenti:2015aaa}
Vitenti S D~P and Penna-Lima M
  \href{https://doi.org/10.1088/1475-7516/2015/9/045}{2015 {\em JCAP} {\bf 09}
  045}

\bibitem{Zhang:2016tto}
Zhang M-J and Xia J-Q \href{https://doi.org/10.1088/1475-7516/2016/12/005}{2016
  {\em JCAP} {\bf 12} 005}

\bibitem{Wei:2016xti}
Wei J-J and Wu X-F \href{https://doi.org/10.3847/1538-4357/aa674b}{2017 {\em
  Astrophys. J.} {\bf 838} 160}

\bibitem{Yu:2017iju}
Yu H, Ratra B, and Wang F-Y
  \href{https://doi.org/10.3847/1538-4357/aab0a2}{2018 {\em Astrophys. J.} {\bf
  856} 3}

\bibitem{Yennapureddy:2017vvb}
Yennapureddy M~K and Melia F
  \href{https://doi.org/10.1088/1475-7516/2017/11/029}{2017 {\em JCAP} {\bf 11}
  029}

\bibitem{Velten:2017ire}
Velten H, Gomes S, and Busti V~C
  \href{https://doi.org/10.1103/PhysRevD.97.083516}{2018 {\em Phys. Rev. D}
  {\bf 97} 083516}

\bibitem{Marra:2017pst}
Marra V and Sapone D \href{https://doi.org/10.1103/PhysRevD.97.083510}{2018
  {\em Phys. Rev. D} {\bf 97} 083510}

\bibitem{Melia:2018tzi}
Melia F and Yennapureddy M~K
  \href{https://doi.org/10.1088/1475-7516/2018/02/034}{2018 {\em JCAP} {\bf 02}
  034}

\bibitem{Gomez-Valent:2018hwc}
G{\'o}mez-Valent A and Amendola L
  \href{https://doi.org/10.1088/1475-7516/2018/04/051}{2018 {\em JCAP} {\bf 04}
  051}

\bibitem{Haridasu:2018gqm}
Haridasu B~S, Lukovi{\'c} V~V, Moresco M, and Vittorio N
  \href{https://doi.org/10.1088/1475-7516/2018/10/015}{2018 {\em JCAP} {\bf 10}
  015}

\bibitem{Capozziello:2018jya}
Capozziello S, Ruchika, and Sen A~A
  \href{https://doi.org/10.1093/mnras/stz176}{2019 {\em Mon. Not. Roy. Astron.
  Soc.} {\bf 484} 4484}

\bibitem{Arjona:2019fwb}
Arjona R and Nesseris S \href{https://doi.org/10.1103/PhysRevD.101.123525}{2020
  {\em Phys. Rev. D} {\bf 101} 123525}

\bibitem{Jesus:2019nnk}
Jesus J~F, Valentim R, Escobal A~A, and Pereira S~H
  \href{https://doi.org/10.1088/1475-7516/2020/04/053}{2020 {\em JCAP} {\bf 04}
  053}

\bibitem{Franco:2019wbj}
Franco F~O, Bonvin C, and Clarkson C
  \href{https://doi.org/10.1093/mnrasl/slz175}{2020 {\em Mon. Not. Roy. Astron.
  Soc.} {\bf 492} L34}

\bibitem{Dhawan:2021mel}
Dhawan S, Alsing J, and Vagnozzi S
  \href{https://doi.org/10.1093/mnrasl/slab058}{2021 {\em Mon. Not. Roy.
  Astron. Soc.} {\bf 506} L1}

\bibitem{Gangopadhyay:2023nli}
Gangopadhyay M~R, Sami M, and Sharma M~K
  \href{https://doi.org/10.1103/PhysRevD.108.103526}{2023 {\em Phys. Rev. D}
  {\bf 108} 103526}

\bibitem{Sharma:2024mtq}
Sharma M~K and Sami M \href{https://doi.org/10.1088/1475-7516/2025/05/002}{2025
  {\em JCAP} {\bf 05} 002}

\bibitem{Dinda:2024kjf}
Dinda B~R \href{https://doi.org/10.1088/1475-7516/2024/09/062}{2024 {\em JCAP}
  {\bf 09} 062}

\bibitem{Jiang:2024xnu}
Jiang J-Q, Pedrotti D, da~Costa S~S, and Vagnozzi S
  \href{https://doi.org/10.1103/PhysRevD.110.123519}{2024 {\em Phys. Rev. D}
  {\bf 110} 123519}

\bibitem{Ghosh:2024kyd}
Ghosh B and Bengaly C \href{https://doi.org/10.1016/j.dark.2024.101699}{2024
  {\em Phys. Dark Univ.} {\bf 46} 101699}

\bibitem{Cortes:2024lgw}
Cort{\^e}s M and Liddle A~R
  \href{https://doi.org/10.1088/1475-7516/2024/12/007}{2024 {\em JCAP} {\bf 12}
  007}

\bibitem{Shlivko:2024llw}
Shlivko D and Steinhardt P~J
  \href{https://doi.org/10.1016/j.physletb.2024.138826}{2024 {\em Phys. Lett.
  B} {\bf 855} 138826}

\bibitem{deCruzPerez:2024shj}
de~Cruz~Perez J, Park C-G, and Ratra B
  \href{https://doi.org/10.1103/PhysRevD.110.023506}{2024 {\em Phys. Rev. D}
  {\bf 110} 023506}

\bibitem{Roy:2024kni}
Roy N \href{https://doi.org/10.1016/j.dark.2025.101912}{2025 {\em Phys. Dark
  Univ.} {\bf 48} 101912}

\bibitem{Chatrchyan:2024xjj}
Chatrchyan A, Niedermann F, Poulin V, and Sloth M~S
  \href{https://doi.org/10.1103/PhysRevD.111.043536}{2025 {\em Phys. Rev. D}
  {\bf 111} 043536}

\bibitem{Perivolaropoulos:2024yxv}
Perivolaropoulos L \href{https://doi.org/10.1103/PhysRevD.110.123518}{2024 {\em
  Phys. Rev. D} {\bf 110} 123518}

\bibitem{Payeur:2024dnq}
Payeur G, McDonough E, and Brandenberger R
  \href{https://doi.org/10.1103/bggr-61nr}{2025 {\em Phys. Rev. D} {\bf 111}
  123541}

\bibitem{Chan-GyungPark:2024brx}
Park C-G, de~Cruz~P{\'e}rez J, and Ratra B
  \href{https://doi.org/10.1142/S0218271825500580}{2025 {\em Int. J. Mod. Phys.
  D} {\bf 34} 2550058}

\bibitem{Gao:2024ily}
Gao Q, Peng Z, Gao S, and Gong Y
  \href{https://doi.org/10.3390/universe11010010}{2025 {\em Universe} {\bf 11}
  10}

\bibitem{Dinda:2024ktd}
Dinda B~R and Maartens R
  \href{https://doi.org/10.1088/1475-7516/2025/01/120}{2025 {\em JCAP} {\bf 01}
  120}

\bibitem{Dinda:2025iaq}
Dinda B~R and Maartens R \href{https://doi.org/10.1093/mnrasl/slaf063}{2025
  {\em Mon. Not. Roy. Astron. Soc.} {\bf 542} L31}

\bibitem{Chen:2025mlf}
Chen S-F and Zaldarriaga M
  \href{https://doi.org/10.1088/1475-7516/2025/08/014}{2025 {\em JCAP} {\bf 08}
  014}

\bibitem{Dinda:2025hiu}
Dinda B~R, Maartens R, and Clarkson C
  \href{https://doi.org/10.1088/1475-7516/2025/12/025}{2025 {\em JCAP} {\bf 12}
  025}

\end{thebibliography}

\end{multicols}

\vskip 0.5cm

\textit{\large Appendix: Joint distribution of the multivariate normal distribution }
\vskip 0.5cm
For the joint distribution,
\begin{equation}
\begin{split}
\begin{bmatrix}
Y_1 \\
Y_2 \\
F^{1}_*
\end{bmatrix} &\sim
\mathcal{N} \left(
\begin{bmatrix}
\mu_1\\
\mu_2\\
\mu_{1*}
\end{bmatrix},
\begin{bmatrix}
K^a(X_1,X_1)+C_{11} & K^{ab}(X_1,X_2)+C_{12} & K^a(X_1,X_{1*}) \\
K^{ba}(X_2,X_1)+C_{12}^T & K^b(X_2,X_2)+C_{22} & K^{ba}(X_2,X_{1*}) \\
K^{a}(X_{1*},X_1) & K^{ab}(X_{1*},X_2) & K^a(X_{1*},X_{1*})
\end{bmatrix}
\right)\\
&=\mathcal{N} \left(
\begin{bmatrix}
\mu_1\\
\mu_2\\
\mu_{1*}
\end{bmatrix},
\begin{bmatrix}
K^a_{1,1}+C_{11} & K^{ab}_{1,2}+C_{12} & K^a_{1,1*} \\
K^{ba}_{2,1}+C_{12}^T & K^b_{2,2}+C_{22} & K^{ba}_{2,1*} \\
K^{a}_{1*,1} & K^{ab}_{1*,2} & K^a_{1*,1*}
\end{bmatrix}
\right),
\end{split}
\end{equation}
we take
\begin{equation}
\tilde{Y}_1 =
\begin{bmatrix}
 Y_{1} \\
 Y_{2}
\end{bmatrix},
\quad
\tilde{\mu}_1 =
\begin{bmatrix}
 \mu_1\\
 \mu_2
\end{bmatrix},
\end{equation}

\begin{equation}
\tilde{\Sigma}_{11} =
\begin{bmatrix}
 K^a_{1,1}+C_{11} & K_{1,2}^{ab}+C_{12}\\
 K_{2,1}^{ba}+C_{12}^T & K_{2,2}^{b}+C_{22}
\end{bmatrix},
\end{equation}
and
\begin{equation}
\tilde{K}_{*1}^1 =
\begin{bmatrix}
 K_{1*,1}^a & K_{1*,2}^{ab}
\end{bmatrix},
\end{equation}
then we get
\begin{equation}
\bar{F}^1_{*} = \mu_{1*} + \tilde{K}_{*1}^1 \tilde{\Sigma}_{11}^{-1} (\tilde{Y}_1-\tilde{\mu}_1)
= \mu_{1*}+K^{a}_{1*,1}B_1+K^{ab}_{1*,2}B_2,
\end{equation}
and
\begin{equation}
\begin{split}
\text{Cov}[F^1_{*},F^1_{*}] &= K^a_{1*,1*} - \tilde{K}^1_{*1} \tilde{\Sigma}_{11}^{-1} \tilde{K}^1_{1*} \\ 
&=K^a_{1*,1*}-K^a_{1*,1}(A_{11}K^a_{1,1*}+A_{12}K_{2,1*}^{ba})-K^{ab}_{1*,2}(A_{21}K^a_{1,1*}+A_{22}K_{2,1*}^{ba}),
\end{split}
\end{equation}
where $C_{11}$ and $C_{22}$ are the measured covariance in data $Y_1$ and $Y_2$, respectively, and $C_{12}$ is the measured covariance between $Y_1$ and $Y_2$, 
the inverse matrix
\begin{equation}
\tilde{\Sigma}_{11}^{-1} =
\begin{bmatrix}
 A_{11} & A_{12}\\
 A_{21} & A_{22}
\end{bmatrix},
\end{equation}
\begin{equation}
B_1 = A_{11}(Y_1-\mu_1)+A_{12}(Y_2-\mu_2),
\end{equation}
and
\begin{equation}
B_2 = A_{21}(Y_1-\mu_1)+A_{22}(Y_2-\mu_2).
\end{equation}
In the linear model of coregionalization (LMC), the covariance structure between the outputs  $Y_1$  and  $Y_2$ is represented as a linear combination of a finite number of shared latent kernels. Specifically, the auto- and cross-covariance blocks are written as:
\begin{gather}
K^a = c_{a1}^2 K^1 + c_{a2}^2 K^2 + c_{a3}^2 K^3,\\
K^b = c_{b1}^2 K^1 + c_{b2}^2 K^2 + c_{b3}^2 K^3,\\
K^{ab} = c_{a1}c_{b1} K^1 + c_{a2}c_{b2} K^2 + c_{a3}c_{b3} K^3,
\end{gather}
where each  $K^i(\cdot,\cdot)$ corresponds to an independent latent Gaussian process, and the coefficients  $c_{ai}$, $c_{bi}$
determine how strongly each latent process contributes to the observed outputs  $Y_1$ and $Y_2$.
The shared latent kernels  $K^i$
 capture the common structure across outputs, while the coefficients control task-specific correlations and relative amplitudes.

The hyperparameters in the kernels $K^a$, $K^{ab}$ and $K^b$ can be obtained by maximizing the likelihood,
\begin{equation}
  -\ln P = \frac{1}{2}(\tilde{Y}_1-\tilde{\mu}_1)^T\tilde{\Sigma}_{11}^{-1} (\tilde{Y}_1-\tilde{\mu}_1) + \frac{1}{2}\ln |\tilde{\Sigma}_{11}| + \frac{N}{2}\ln(2\pi).
\end{equation}

Similarly, for the joint distribution,
\begin{equation}
\begin{bmatrix}
Y_1 \\
Y_2 \\
F^{2}_*
\end{bmatrix} \sim
\mathcal{N} \left(
\begin{bmatrix}
\mu_1\\
\mu_2\\
\mu_{2*}
\end{bmatrix},
\begin{bmatrix}
K^a_{1,1}+C_{11} & K^{ab}_{1,2}+C_{12} & K^{ab}_{1,2*} \\
K^{ba}_{2,1}+C_{12}^T & K^b_{2,2}+C_{22} & K^{b}_{2,2*} \\
K^{ba}_{2*,1} & K^{b}_{2*,2} & K^b_{2*,2*}
\end{bmatrix}
\right),
\end{equation}
we get
\begin{equation}
\tilde{K}_{*1}^2 =
\begin{bmatrix}
 K_{2*,1}^{ba} & K_{2*,2}^{b}
\end{bmatrix},
\end{equation}
\begin{equation}
\bar{F}^2_{*} = \mu_{2*} + \tilde{K}_{*1}^2 \tilde{\Sigma}_{11}^{-1} (\tilde{Y}_1-\tilde{\mu}_1)
= \mu_{2*}+K^{ba}_{2*,1}B_1+K^{b}_{2*,2}B_2,
\end{equation}
and
\begin{equation}
\begin{split}
\text{Cov}[F^2_{*},F^2_{*}] &= K^b_{2*,2*} - \tilde{K}^2_{*1} \tilde{\Sigma}_{11}^{-1} \tilde{K}^2_{1*} \\ 
&=K^b_{2*,2*}-K^{ba}_{2*,1}(A_{11}K^{ab}_{1,2*}+A_{12}K_{2,2*}^{b})
-K^{b}_{2*,2}(A_{21}K^{ab}_{1,2*}+A_{22}K_{2,2*}^{b}).
\end{split}
\end{equation}

For the same reason, we can derive all the mean and covariance matrix,
\begin{equation}
\bar{F^1}'_* = \mu'_{1*} + \tilde{K}_{*1}^{1(1,0)}\tilde{\Sigma}_{11}^{-1} (\tilde{Y}_1-\tilde{\mu}_1) 
= \mu'_{1*}+K_{1*,1}^{a(1,0)}B_1+K_{1*,2}^{ab(1,0)}B_2,
\end{equation}
\begin{equation}
\bar{F^1}''_* = \mu''_{1*} + \tilde{K}_{*1}^{1(2,0)}\tilde{\Sigma}_{11}^{-1} (\tilde{Y}_1-\tilde{\mu}_1) 
= \mu''_{1*}+K_{1*,1}^{a(2,0)}B_1+K_{1*,2}^{ab(2,0)}B_2,
\end{equation}
\begin{equation}
\bar{F^2}'_* = \mu'_{2*} + \tilde{K}_{*1}^{2(1,0)}\tilde{\Sigma}_{11}^{-1} (\tilde{Y}_1-\tilde{\mu}_1)
= \mu'_{2*}+K_{2*,1}^{ba(1,0)}B_1+K_{2*,2}^{b(1,0)}B_2,
\end{equation}
\begin{equation}
\bar{F^2}''_* = \mu''_{2*} + \tilde{K}_{*1}^{2(2,0)}\tilde{\Sigma}_{11}^{-1} (\tilde{Y}_1-\tilde{\mu}_1)
= \mu''_{2*}+K_{2*,1}^{ba(2,0)}B_1+K_{2*,2}^{b(2,0)}B_2,
\end{equation}

\begin{equation}
\begin{split}
\text{Cov}[F^{1\prime}_*,F^{1\prime}_*] &= K_{1*,1*}^{a(1,1)} - \tilde{K}_{*1}^{1(1,0)} \tilde{\Sigma}_{11}^{-1} \tilde{K}_{1*}^{1(0,1)} \\ 
&=K_{1*,1*}^{a(1,1)}-K_{1*,1}^{a(1,0)}(A_{11}K_{1,1*}^{a(0,1)}+A_{12}K_{2,1*}^{ba(0,1)})
-K_{1*,2}^{ab(1,0)}(A_{21}K_{1,1*}^{a(0,1)}+A_{22}K_{2,1*}^{ba(0,1)}).
\end{split}
\end{equation}
\begin{equation}
\begin{split}
\text{Cov}[{F^1}''_*,{F^1}''_*] &= K_{1*,1*}^{a(2,2)} - \tilde{K}_{*1}^{1(2,0)} \tilde{\Sigma}_{11}^{-1} \tilde{K}_{1*}^{1(0,2)} \\ &=K_{1*,1*}^{a(2,2)}-K_{1*,1}^{a(2,0)}(A_{11}K_{1,1*}^{a(0,2)}+A_{12}K_{2,1*}^{ba(0,2)})
-K_{1*,2}^{ab(2,0)}(A_{21}K_{1,1*}^{a(0,2)}+A_{22}K_{2,1*}^{ba(0,2)}).
\end{split}
\end{equation}
\begin{equation}
\begin{split}
\text{Cov}[F^{2\prime}_*,F^{2\prime}_*] &= K_{2*,2*}^{b(1,1)} - \tilde{K}_{*1}^{2(1,0)} \tilde{\Sigma}_{11}^{-1} \tilde{K}_{1*}^{2(0,1)} \\ &=K_{2*,2*}^{b(1,1)}-K_{2*,1}^{ba(1,0)}(A_{11}K_{1,2*}^{ab(0,1)}+A_{12}K_{2,2*}^{b(0,1)})
-K_{2*,2}^{b(1,0)}(A_{21}K_{1,2*}^{ab(0,1)}+A_{22}K_{2,2*}^{b(0,1)}).
\end{split}
\end{equation}
\begin{equation}
\begin{split}
\text{Cov}[{F^2}''_*,{F^2}''_*] &= K_{2*,2*}^{b(2,2)} - \tilde{K}_{*1}^{2(2,0)} \tilde{\Sigma}_{11}^{-1} \tilde{K}_{1*}^{2(0,2)} \\ &=K_{2*,2*}^{b(2,2)}-K_{2*,1}^{ba(2,0)}(A_{11}K_{1,2*}^{ab(0,2)}+A_{12}K_{2,2*}^{b(0,2)})
-K_{2*,2}^{b(2,0)}(A_{21}K_{1,2*}^{ab(0,2)}+A_{22}K_{2,2*}^{b(0,2)}).
\end{split}
\end{equation}
\begin{equation}
\begin{split}
\text{Cov}[F^1_*,{F^1}'_*] &= K_{1*,1*}^{a(0,1)} - \tilde{K}_{*1}^1 \tilde{\Sigma}_{11}^{-1} \tilde{K}_{1*}^{1(0,1)} \\ &=K_{1*,1*}^{a(0,1)}-K_{1*,1}^a(A_{11}K_{1,1*}^{a(0,1)}+A_{12}K_{2,1*}^{ba(0,1)})
-K_{1*,2}^{ab}(A_{21}K_{1,1*}^{a(0,1)}+A_{22}K_{2,1*}^{ba(0,1)}).
\end{split}
\end{equation}
\begin{equation}
\begin{split}
\text{Cov}[F^1_*,{F^1}''_*] &= K_{1*,1*}^{a(0,2)} - \tilde{K}_{*1}^1 \tilde{\Sigma}_{11}^{-1} \tilde{K}_{1*}^{1(0,2)} \\ &=K_{1*,1*}^{a(0,2)}-K_{1*,1}^a(A_{11}K_{1,1*}^{a(0,2)}+A_{12}K_{2,1*}^{ba(0,2)})
-K_{1*,2}^{ab}(A_{21}K_{1,1*}^{a(0,2)}+A_{22}K_{2,1*}^{ba(0,2)}).
\end{split}
\end{equation}
\begin{equation}
\begin{split}
\text{Cov}[{F^1}'_*,{F^1}''_*] &= K_{1*,1*}^{a(1,2)} - \tilde{K}_{*1}^{1(1,0)} \tilde{\Sigma}_{11}^{-1} \tilde{K}_{1*}^{1(0,2)} \\ &=K_{1*,1*}^{a(1,2)}-K_{1*,1}^{a(1,0)}(A_{11}K_{1,1*}^{a(0,2)}+A_{12}K_{2,1*}^{ba(0,2)})
-K_{1*,2}^{ab(1,0)}(A_{21}K_{1,1*}^{a(0,2)}+A_{22}K_{2,1*}^{ba(0,2)}).
\end{split}
\end{equation}
\begin{equation}
\begin{split}
\text{Cov}[F^2_*,{F^2}'_*] &= K_{2*,2*}^{b(0,1)} - \tilde{K}_{*1}^2 \tilde{\Sigma}_{11}^{-1} \tilde{K}_{1*}^{2(0,1)} \\ &=K_{2*,2*}^{b(0,1)}-K_{2*,1}^{ba}(A_{11}K_{1,2*}^{ab(0,1)}+A_{12}K_{2,2*}^{b(0,1)})
-K_{2*,2}^{b}(A_{21}K_{1,2*}^{ab(0,1)}+A_{22}K_{2,2*}^{b(0,1)}).
\end{split}
\end{equation}
\begin{equation}
\begin{split}
\text{Cov}[F^2_*,{F^2}''_*] &= K_{2*,2*}^{b(0,2)} - \tilde{K}_{*1}^2 \tilde{\Sigma}_{11}^{-1} \tilde{K}_{1*}^{2(0,2)} \\ &=K_{2*,2*}^{b(0,2)}-K_{2*,1}^{ba}(A_{11}K_{1,2*}^{ab(0,2)}+A_{12}K_{2,2*}^{b(0,2)})
-K_{2*,2}^{b}(A_{21}K_{1,2*}^{ab(0,2)}+A_{22}K_{2,2*}^{b(0,2)}).
\end{split}
\end{equation}
\begin{equation}
\begin{split}
\text{Cov}[{F^2}'_*,{F^2}''_*] &= K_{2*,2*}^{b(1,2)} - \tilde{K}_{*1}^{2(1,0)} \tilde{\Sigma}_{11}^{-1} \tilde{K}_{1*}^{2(0,2)} \\ &=K_{2*,2*}^{b(1,2)}-K_{2*,1}^{ba(1,0)}(A_{11}K_{1,2*}^{ab(0,2)}+A_{12}K_{2,2*}^{b(0,2)})
-K_{2*,2}^{b(1,0)}(A_{21}K_{1,2*}^{ab(0,2)}+A_{22}K_{2,2*}^{b(0,2)}).
\end{split}
\end{equation}
\begin{equation}
\text{Cov}[F^1_{*},F^2_{*}] =K^{ab}_{1*,2*}-K^a_{1*,1}(A_{11}K^{ab}_{1,2*}+A_{12}K_{2,2*}^{b})
-K^{ab}_{1*,2}(A_{21}K^{ab}_{1,2*}+A_{22}K_{2,2*}^{b}).
\end{equation}
\begin{equation}
\text{Cov}[F^1_{*},{F^2}'_{*}] = K_{1*,2*}^{ab(0,1)}-K_{1*,1}^a(A_{11}K_{1,2*}^{ab(0,1)}+A_{12}K_{2,2*}^{b(0,1)})
-K_{1*,2}^{ab}(A_{21}K_{1,2*}^{ab(0,1)}+A_{22}K_{2,2*}^{b(0,1)}).
\end{equation}
\begin{equation}
\text{Cov}[F^1_{*},{F^2}''_{*}] = K_{1*,2*}^{ab(0,2)}-K_{1*,1}^a(A_{11}K_{1,2*}^{ab(0,2)}+A_{12}K_{2,2*}^{b(0,2)})
-K_{1*,2}^{ab}(A_{21}K_{1,2*}^{ab(0,2)}+A_{22}K_{2,2*}^{b(0,2)}).
\end{equation}
\begin{equation}
\text{Cov}[{F^1}'_{*},{F^2}_{*}] = K_{1*,2*}^{ab(1,0)}-K_{1*,1}^{a(1,0)}(A_{11}K_{1,2*}^{ab}+A_{12}K_{2,2*}^{b})
-K_{1*,2}^{ab(1,0)}(A_{21}K_{1,2*}^{ab}+A_{22}K_{2,2*}^{b}).
\end{equation}
\begin{equation}
\text{Cov}[{F^1}'_{*},{F^2}'_{*}] = K_{1*,2*}^{ab(1,1)}-K_{1*,1}^{a(1,0)}(A_{11}K_{1,2*}^{ab(0,1)}+A_{12}K_{2,2*}^{b(0,1)})
-K_{1*,2}^{ab(1,0)}(A_{21}K_{1,2*}^{ab(0,1)}+A_{22}K_{2,2*}^{b(0,1)}).
\end{equation}
\begin{equation}
\text{Cov}[{F^1}'_{*},{F^2}''_{*}] = K_{1*,2*}^{ab(1,2)}-K_{1*,1}^{a(1,0)}(A_{11}K_{1,2*}^{ab(0,2)}+A_{12}K_{2,2*}^{b(0,2)})
-K_{1*,2}^{ab(1,0)}(A_{21}K_{1,2*}^{ab(0,2)}+A_{22}K_{2,2*}^{b(0,2)}).
\end{equation}
\begin{equation}
\text{Cov}[{F^1}''_{*},{F^2}_{*}] = K_{1*,2*}^{ab(2,0)}-K_{1*,1}^{a(2,0)}(A_{11}K_{1,2*}^{ab}+A_{12}K_{2,2*}^{b})
-K_{1*,2}^{ab(2,0)}(A_{21}K_{1,2*}^{ab}+A_{22}K_{2,2*}^{b}).
\end{equation}
\begin{equation}
\text{Cov}[{F^1}''_{*},{F^2}'_{*}] = K_{1*,2*}^{ab(2,1)}-K_{1*,1}^{a(2,0)}(A_{11}K_{1,2*}^{ab(0,1)}+A_{12}K_{2,2*}^{b(0,1)})
-K_{1*,2}^{ab(2,0)}(A_{21}K_{1,2*}^{ab(0,1)}+A_{22}K_{2,2*}^{b(0,1)}).
\end{equation}
\begin{equation}
\text{Cov}[{F^1}''_{*},{F^2}''_{*}] = K_{1*,2*}^{ab(2,2)}-K_{1*,1}^{a(2,0)}(A_{11}K_{1,2*}^{ab(0,2)}+A_{12}K_{2,2*}^{b(0,2)})
-K_{1*,2}^{ab(2,0)}(A_{21}K_{1,2*}^{ab(0,2)}+A_{22}K_{2,2*}^{b(0,2)}).
\end{equation}

\end{document}